\documentstyle[twocolumn,prl,aps,epsf,axodraw]{revtex}

\def\text{\textstyle}

\newcommand{\lsim}{\raisebox{-0.13cm}{~\shortstack{$<$ \\[-0.07cm] $\sim$}}~}
\newcommand{\gsim}{\raisebox{-0.13cm}{~\shortstack{$>$ \\[-0.07cm] $\sim$}}~}

\newcommand{\TeV}{\unskip\,\mathrm{TeV}}
\newcommand{\GeV}{\unskip\,\mathrm{GeV}}
\newcommand{\MeV}{\unskip\,\mathrm{MeV}}

\newcommand{\fb}{\unskip\,\mathrm{fb}}

\def\draftdate{\relax}
\def\mda{\relax}
\def\mua{\relax}
\def\mla{\relax}
\def\draft{
\def\thtystars{******************************}
\def\sixtystars{\thtystars\thtystars}
\typeout{}
\typeout{\sixtystars**}
\typeout{* Draft mode!
         For final version remove \protect\draft\space in source file *}
\typeout{\sixtystars**}
\typeout{}
\def\draftdate{\today}
\def\mua{\marginpar[\boldmath\hfil$\uparrow$]%
                   {\boldmath$\uparrow$\hfil}%
                    \typeout{marginpar: $\uparrow$}\ignorespaces}
\def\mda{\marginpar[\boldmath\hfil$\downarrow$]%
                   {\boldmath$\downarrow$\hfil}%
                    \typeout{marginpar: $\downarrow$}\ignorespaces}
\def\mla{\marginpar[\boldmath\hfil$\rightarrow$]%
                   {\boldmath$\leftarrow $\hfil}%
                    \typeout{marginpar: $\leftrightarrow$}\ignorespaces}
\overfullrule 5pt
\oddsidemargin -15mm
\marginparwidth 29mm
}

\catcode`\@=11
\ifx\@Hxfloat\@Hundef\else\expandafter\endinput\fi
\let\@Hxfloat\@xfloat
\def\@xfloat#1[{\@ifnextchar{H}{\@HHfloat{#1}[}{\@Hxfloat{#1}[}}
\def\@HHfloat#1[H]{%
\expandafter\let\csname end#1\endcsname\end@Hfloat
\vskip\intextsep\vbox\bgroup\def\@captype{#1}\parindent\z@
\ignorespaces}
\def\end@Hfloat{\egroup\vskip \intextsep}
\def\section{\@startsection{section}{1}{\z@}{3.5ex plus 1ex minus .2ex}
{2.3ex plus .2ex}{\large\bf}}
\def\thesection{\arabic{section}.}
\def\thesubsection{\arabic{section}.\arabic{subsection}.}

\def\appendix{\setcounter{section}{0}
 \def\thesection{APPENDIX \Alph{section}:}
 \def\theequation{\Alph{section}.\arabic{equation}}}
\def\@citex[#1]#2{\if@filesw\immediate\write\@auxout{\string\citation{#2}}\fi
  \def\@citea{}\@cite{\@for\@citeb:=#2\do
    {\@citea\def\@citea{,\penalty\@m}\@ifundefined
       {b@\@citeb}{{\bf ?}\@warning
       {Citation `\@citeb' on page \thepage \space undefined}}%
\hbox{\csname b@\@citeb\endcsname}}}{#1}}

\def\citer{\@ifnextchar [{\@tempswatrue\@citexr}{\@tempswafalse\@citexr[]}}
\def\@citexr[#1]#2{\if@filesw\immediate\write\@auxout{\string\citation{#2}}\fi
  \def\@citea{}\@cite{\@for\@citeb:=#2\do
    {\@citea\def\@citea{--\penalty\@m}\@ifundefined
       {b@\@citeb}{{\bf ?}\@warning
       {Citation `\@citeb' on page \thepage \space undefined}}%
\hbox{\csname b@\@citeb\endcsname}}}{#1}}
\begin{document}

\renewcommand{\thefootnote}{\fnsymbol{footnote}}
\setcounter{page}{0}
\pagestyle{empty}
\onecolumn

\begin{large}

\begin{flushright}
DESY 2001--077 \\
Edinburgh 2001/08 \\
PSI--PR--01--10 \\
hep-ph/0107081
\end{flushright}

\vspace*{1cm}

\begin{center}

{\Large \bf \sc Higgs Radiation off Top Quarks \\[.5em]
at the Tevatron and the LHC}

\vspace*{1cm}

{\sc W.~Beenakker$^1$%
\footnote{Research supported in part by a PPARC Research Fellowship.},
S.~Dittmaier$^2$%
\footnote{Heisenberg Fellow of Deutsche Forschungsgemeinschaft DFG, Bonn.},
M.~Kr\"amer$^3$, B.~Pl\"umper$^2$, \\[.3em]
M.~Spira$^4$ and P.M.~Zerwas$^2$}

\vspace*{1cm}

{\normalsize \it
$^1$ Theoretical Physics, University of Nijmegen,
NL-6500 GL Nijmegen, The Netherlands\\[.2cm]
$^2$ Deutsches Elektronen-Synchrotron DESY, D-22603 Hamburg, Germany \\[.2cm]
$^3$ Department~of Physics and Astronomy, University~of~Edinburgh,
Edinburgh EH9 3JZ, Scotland \\[.2cm]
$^4$ Paul Scherrer Institut PSI, CH-5232 Villigen PSI, Switzerland}
\\[2cm]

{\normalsize \bf Abstract} \\[.5cm]

\begin{minipage}{16.0cm}
\noindent {\normalsize
Higgs bosons can be searched for in the channels $p\bar p/pp\to t\bar
tH+X$ at the Tevatron and the LHC. We have calculated the QCD
corrections to these processes in the Standard Model at
next-to-leading order.  The higher-order corrections reduce the
renormalization and factorization scale dependence considerably and
%PPPP
stabilize the theoretical predictions for the cross sections.  At the
central
scale $\mu=(2m_t+M_H)/2$ the properly defined $K$ factors are slightly below
unity for the Tevatron ($K \sim 0.8$) and slightly above
unity for the LHC ($K \sim 1.2$).}
\end{minipage}

\end{center}
\end{large}

\twocolumn
\pagestyle{plain}

\title{\bf \large \sc Higgs Radiation off Top Quarks %\\[.5em]
at the Tevatron and the LHC}

\author{\sc W.~Beenakker$^1$, S.~Dittmaier$^2$,
M.~Kr\"amer$^3$, B.~Pl\"umper$^2$, %\\[.3em]
M.~Spira$^4$ and P.M.~Zerwas$^2$}

\address{
$^1$ Theoretical Physics, University of Nijmegen, P.O.\ Box 9010,
NL-6500 GL Nijmegen, The Netherlands \\
$^2$ Deutsches Elektronen-Synchrotron DESY, D-22603 Hamburg, Germany \\
$^3$ Department~of Physics and Astronomy, University~of~Edinburgh,
Edinburgh EH9 3JZ, Scotland \\
$^4$ Paul Scherrer Institut PSI, CH-5232 Villigen PSI, Switzerland
\\[0.5cm]
%\abstract{
\begin{minipage}{16.0cm}
\noindent \rm
Higgs bosons can be searched for in the channels $p\bar p/pp\to t\bar
tH+X$ at the Tevatron and the LHC. We have calculated the QCD
corrections to these processes in the Standard Model at
next-to-leading order.  The higher-order corrections reduce the
renormalization and factorization scale dependence considerably and
stabilize the theoretical predictions for the cross sections. At the
central
scale $\mu=(2m_t+M_H)/2$ the properly defined $K$ factors are slightly below
unity for the Tevatron ($K \sim 0.8$) and slightly above
unity for the LHC ($K \sim 1.2$).
\end{minipage}
}
%\date{\today}
\maketitle

%\pacs{12.60.Jv, 14.80.Cp, 12.38.Bx}

\begin{narrowtext}
\renewcommand{\thefootnote}{\arabic{footnote}}

\setcounter{footnote}{0}
\vskip-1.43cm

\noindent
{\bf 1.} The search for Higgs bosons \cite{Higgs:1964ia} is one of the
most important experimental programs in high-energy physics. If
successful, a crucial step in revealing the mechanism for electroweak
symmetry breaking and the generation of masses for the fundamental
particles in the Standard Model (SM), electroweak gauge bosons,
leptons, and quarks, will have been taken. In the near future, the
search for Higgs bosons will be carried out at hadron colliders, the
proton--antiproton collider Tevatron \cite{Carena:2000yx} with a
center-of-mass (CM) energy of $2\TeV$, followed by the proton--proton
collider LHC~\cite{atlas_cms_tdrs} with $14\TeV$. Analyses of
precision electroweak data \cite{:2001xv} set the focus on $M_H\lsim
200\GeV$ as the preferential Higgs mass range in the SM, although a firm
prediction without escape roads is not possible
\cite{Chanowitz:2001bv}.

Various channels can be exploited at hadron colliders to search for a
Higgs boson in the intermediate mass range. Among these channels Higgs
radiation off top quarks \cite{Kunszt:1984ri} plays an important
r\^ole:
\begin{equation}
p\bar p / pp \to t\bar tH\;+\;X \qquad \mbox{via} \qquad
q\bar q, gg\to t\bar tH.
\label{eq:procs}
\end{equation}

Although the expected rate is low at the Tevatron, a sample of a few
but very clean events could be observed for Higgs masses below
$140\GeV$, while this channel becomes very demanding above
\cite{Goldstein:2001bp}. At the LHC associated production of the Higgs
boson with top quarks is an important search channel for Higgs masses
below $\sim 125\GeV$.
%, i.e.\ below the region where the $ZZ^*$ final states
%become important.
Moreover, analyzing the $t\bar tH$ production rate at the LHC can
provide information on the top--Higgs Yukawa coupling, assuming
standard decay branching ratios \cite{Drollinger:2001xm}, before model
independent precision measurements of this coupling will be performed
at $e^+e^-$ colliders \cite{Djouadi:1992tk}.

\vspace*{0.5em}
\noindent
{\bf 2.} Predictions for the cross sections (\ref{eq:procs}), which
are based on the leading order (LO), are plagued by considerable
uncertainties due to the strong dependence on the renormalization and
factorization scales, introduced by the QCD coupling and the parton
densities (see the figures below).  While estimates of radiative
corrections have been presented before in the ``effective Higgs
approximation'' (EHA) \cite{Dawson:1998im}, in this letter we present
the first complete calculation of the QCD corrections at
next-to-leading order (NLO), which reduce the spurious scale
dependence significantly and lead to stable predictions for the cross
sections.%
%\footnote{Each part of the calculation has been worked out twice and
%by two independent parts of the collaboration.}

The Born diagrams, generic examples of which are displayed in
Fig.~\ref{fig:LOdiags}(a), are supplemented in NLO by virtual
gluon-exchange diagrams, Fig.~\ref{fig:LOdiags}(b), running in
complexity up to pentagons; by gluon radiation,
Fig.~\ref{fig:LOdiags}(c); and by parton splitting,
Fig.~\ref{fig:LOdiags}(d). The latter two add incoherently to the
virtual corrections.
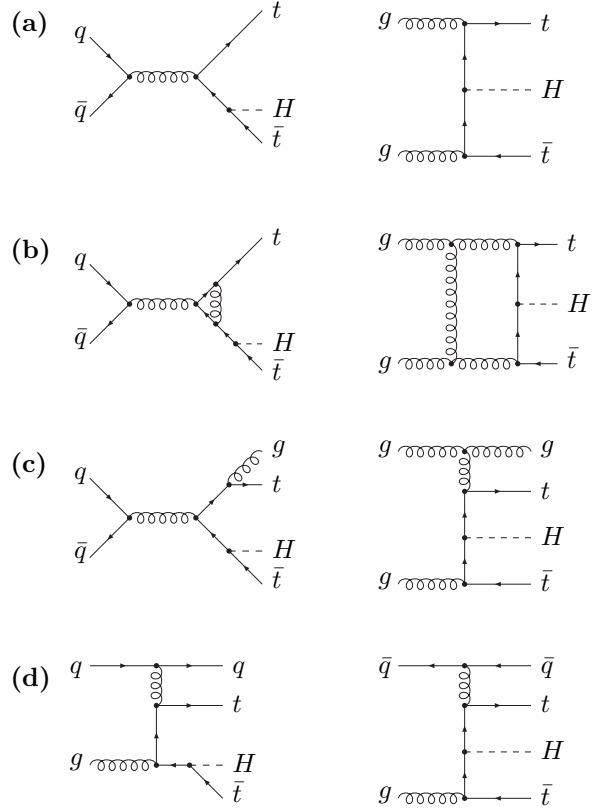
\begin{figure}[hbt]
\SetScale{0.5}
\noindent
{\unitlength 0.5pt
\begin{picture}(170,120)(-50,-10)
\ArrowLine(20, 80)(50,50)
\ArrowLine(50, 50)(20,20)
\Gluon(50,50)(100,50){4}{5}
\Vertex(50,50){2}
\Vertex(100,50){2}
\Vertex(125,25){2}
\DashLine(125, 25)(150,25){5}
\ArrowLine(100,50)(150,100)
\ArrowLine(125,25)(100, 50)
\ArrowLine(150, 0)(125, 25)
\put( 8,80){$q$}
\put( 8,20){$\bar q$}
\put(158, 95){$t$}
\put(158,20){$H$}
\put(158,-3){$\bar{t}$}
\put(-40, 85){{\bf (a)}}
\end{picture}
\hspace*{1em}
\begin{picture}(170,120)(-50,0)
\Gluon(50,100)(100,100){4}{5}
\Gluon(50,  0)(100,  0){4}{5}
\Vertex(100,100){2}
\Vertex(100,  0){2}
\Vertex(100, 50){2}
\DashLine(100, 50)(150,50){5}
\ArrowLine(100,100)(150,100)
\ArrowLine(100, 50)(100,100)
\ArrowLine(100,  0)(100, 50)
\ArrowLine(150,  0)(100,  0)
\put(158, 95){$t$}
\put(158,44){$H$}
\put(158,-5){$\bar{t}$}
\put(35,100){$g$}
\put(35,  0){$g$}
\end{picture}
}
\\[2em]
{\unitlength 0.5pt
\begin{picture}(170,120)(-50,0)
\ArrowLine(20, 80)(50,50)
\ArrowLine(50, 50)(20,20)
\Gluon(50,50)(100,50){4}{5}
\Vertex(50,50){2}
\Vertex(100,50){2}
\Vertex(130,20){2}
\Vertex(115,65){2}
\Vertex(115,35){2}
\DashLine(130, 20)(150,20){5}
\ArrowLine(115,65)(150,100)
\ArrowLine(100,50)(115, 65)
\ArrowLine(115,35)(100, 50)
\ArrowLine(130,20)(115,35)
\ArrowLine(150, 0)(130, 20)
\Gluon(115,65)(115,35){4}{3}
\put( 8,80){$q$}
\put( 8,20){$\bar q$}
\put(158, 95){$t$}
\put(158,15){$H$}
\put(158,-7){$\bar{t}$}
\put(-40, 85){{\bf (b)}}
\end{picture}
\hspace*{1em}
\begin{picture}(170,120)(-50,0)
\Gluon(50, 95)( 90, 95){4}{4}
\Gluon(50,  5)( 90,  5){4}{4}
\Gluon(90, 95)(140, 95){4}{5}
\Gluon(90,  5)(140,  5){4}{5}
\Gluon(90, 95)( 90,  5){4}{9}
\Vertex( 90, 95){2}
\Vertex( 90,  5){2}
\Vertex(140,  5){2}
\Vertex(140, 50){2}
\Vertex(140, 95){2}
\DashLine(140,50)(170,50){5}
\ArrowLine(140, 95)(170, 95)
\ArrowLine(140, 50)(140, 95)
\ArrowLine(140,  5)(140, 50)
\ArrowLine(170,  5)(140,  5)
\put(178, 90){$t$}
\put(178,44){$H$}
\put(178, 2){$\bar{t}$}
\put(35, 95){$g$}
\put(35,   2){$g$}
\end{picture}
}
\\[2em]
{\unitlength 0.5pt
\begin{picture}(170,120)(-50,0)
\ArrowLine(20, 80)(50,50)
\ArrowLine(50, 50)(20,20)
\Gluon(50,50)(100,50){4}{5}
\Vertex(50,50){2}
\Vertex(100,50){2}
\Vertex(125,25){2}
\Vertex(125,75){2}
\DashLine(125, 25)(150,25){5}
\ArrowLine(125,75)(150, 75)
\ArrowLine(100,50)(125, 75)
\ArrowLine(125,25)(100, 50)
\ArrowLine(150, 0)(125, 25)
\Gluon(125,75)(150,100){4}{3}
\put(158,100){$g$}
\put( 8,80){$q$}
\put( 8,20){$\bar q$}
\put(158, 70){$t$}
\put(158,20){$H$}
\put(158,-3){$\bar{t}$}
\put(-40, 85){\bf (c)}
\end{picture}
\hspace*{1em}
\begin{picture}(170,120)(-50,0)
\Gluon(50,100)(100,100){4}{5}
\Gluon(100,100)(150,100){4}{5}
\Gluon(100,100)(100,70){4}{3}
\Gluon(50,  0)(100,  0){4}{5}
\Vertex(100,100){2}
\Vertex(100, 70){2}
\Vertex(100,  0){2}
\Vertex(100, 35){2}
\DashLine(100, 35)(150,35){5}
\ArrowLine(100, 70)(150, 70)
\ArrowLine(100, 35)(100, 70)
\ArrowLine(100,  0)(100, 35)
\ArrowLine(150,  0)(100,  0)
\put(158,100){$g$}
\put(158, 65){$t$}
\put(158,30){$H$}
\put(158,-5){$\bar{t}$}
\put(35,100){$g$}
\put(35,  0){$g$}
\end{picture}
}
\\[2em]
{\unitlength 0.5pt
\begin{picture}(170,120)(-20,0)
\ArrowLine(50,100)(100,100)
\ArrowLine(100,100)(150,100)
\Gluon(100,100)(100,70){4}{3}
\Gluon(50, 25)(100, 25){4}{5}
\Vertex(100,100){2}
\Vertex(100, 70){2}
\Vertex(100, 25){2}
\Vertex(125, 25){2}
\DashLine(125, 25)(150, 25){5}
\ArrowLine(100, 70)(150, 70)
\ArrowLine(100, 25)(100, 70)
\ArrowLine(125, 25)(100, 25)
\ArrowLine(150, 0)(125, 25)
\put(158, 95){$q$}
\put(158, 65){$t$}
\put(158,20){$H$}
\put(158,-3){$\bar{t}$}
\put(35, 95){$q$}
\put(35, 25){$g$}
\put(-10, 85){\bf (d)}
\end{picture}
\hspace*{1em}
\begin{picture}(170,120)(-50,0)
\ArrowLine(100,100)(50,100)
\ArrowLine(150,100)(100,100)
\Gluon(100,100)(100,70){4}{3}
\Gluon(50,  0)(100,  0){4}{5}
\Vertex(100,100){2}
\Vertex(100, 70){2}
\Vertex(100,  0){2}
\Vertex(100, 35){2}
\DashLine(100, 35)(150,35){5}
\ArrowLine(100, 70)(150, 70)
\ArrowLine(100, 35)(100, 70)
\ArrowLine(100,  0)(100, 35)
\ArrowLine(150,  0)(100,  0)
\put(158, 65){$t$}
\put(158,30){$H$}
\put(158,-5){$\bar{t}$}
\put(35,  0){$g$}
\put(158, 95){$\bar q$}
\put(35, 95){$\bar q$}
\end{picture}
}
\vspace*{3em}
\caption{A generic set of diagrams (a) for the Born level,
(b) for virtual gluon exchange, (c) gluon radiation and (d) parton
splitting in the subprocesses $q\bar q, gg\to t\bar tH$ etc.}
\label{fig:LOdiags}
\end{figure}

Dimensional regularization has been adopted for isolating the
ultraviolet, infrared and collinear singularities. Renormalization and
factorization are performed in the $\overline{\mathrm{MS}}$ scheme
with the top mass defined on-shell. The top quark is decoupled from
the running of the strong coupling $\alpha_{\mathrm{s}}(\mu)$.  For
the evaluation of the $p\bar p/pp$ cross sections we have adopted the
CTEQ4L and CTEQ4M \cite{Lai:1997mg} parton densities at LO and NLO,
corresponding to the QCD parameters $\Lambda_5^{\mathrm{LO}}=181\MeV$
and $\Lambda_5^{\overline{\mathrm{MS}}}=202\MeV$ at the one- and
two-loop level of $\alpha_{\mathrm{s}}(\mu)$, respectively.  The
strength of the SM Yukawa coupling is fixed by $g_{ttH}=m_t/v$, where
$v=246\GeV$ is the vacuum-expectation value of the Higgs field, and
the top-quark mass is set to $m_t=174\GeV$.

The most complicated one-loop diagrams are the pentagons, both
analytically and numerically. To calculate a five-point integral
$E^{(D)}$ in $D$ dimensions, the singularity structure
$E_{\mathrm{sing}}^{(D)}$ in $D$ dimensions is determined first.  The
singular part $E_{\mathrm{sing}}^{(D)}$ is given entirely by
three-point subintegrals.  The difference
$E^{(D)}-E_{\mathrm{sing}}^{(D)}$ is finite and regularization-scheme
independent. Therefore it can be calculated in the convenient mass
regularization scheme in 4 dimensions. The original integral $E^{(D)}$
then reads
\begin{eqnarray}
E^{(D)} &=& E_{\mathrm{sing}}^{(D)} +
\left[E^{\mathrm{(mass;D=4)}}-E_{\mathrm{sing}}^{\mathrm{(mass;D=4)}}
\right]
\label{eq:pentagon}
\end{eqnarray}
in the limits $D \to 4$ and $mass \to 0$.
Since $E^{\mathrm{(mass;D=4)}}$ can be expressed in terms of
four-point functions \cite{Melrose:1965kb}, the $D$-dimensional
five-point integral $E^{(D)}$ is finally reduced to three- and
four-point functions.  These integrals and their tensor structures can
be treated according to standard methods \cite{'tHooft:1979xw}.
Numerical instabilities caused by vanishing Gram determinants near the
phase-space boundary, can be controlled by careful extrapolation out
of the safe inner phase-space domains.
[Technical details will be presented in a subsequent publication.]

To extract the singularities of the real part of the NLO corrections
$\sigma^{\mathrm{real}}$, a generalization of the dipole subtraction
formalism \cite{Catani:1996jh} to massive quarks \cite{Catani:2001}
has been adopted (see also Ref.~\cite{Dittmaier:2000mb}).  The
singularities of the cross section $\sigma^{\mathrm{real}}$ are mapped
onto a suitably chosen auxiliary cross section $\sigma^{\mathrm{sub}}$
which is still simple enough so that the singular regions in phase
space can be integrated out analytically, while the difference
$\sigma^{\mathrm{real}}-\sigma^{\mathrm{sub}}$ can safely be
integrated numerically in 4 dimensions.  The auxiliary cross section
$\sigma^{\mathrm{sub}}$ can be decomposed into a part
$\sigma^{\mathrm{sub}}_1$ that, defined on configurations with LO
kinematics, cancels the soft and collinear singularities of the
virtual corrections; and a second part $\sigma^{\mathrm{sub}}_2$
that includes the singularities from initial-state parton splitting,
which are absorbed in the renormalization of the parton
densities. Thus the total NLO correction $\Delta\sigma^{\mathrm{NLO}}$
may be written as the sum
\looseness -1
\begin{eqnarray}
\Delta\sigma^{\mathrm{NLO}} & = &
\left[\sigma^{\mathrm{real}}-\sigma^{\mathrm{sub}}\right]
\;+\; \left[\sigma^{\mathrm{virtual}}+\sigma^{\mathrm{sub}}_1\right]
\nonumber \\
& + &  \left[\sigma^{\mathrm{part}}+\sigma^{\mathrm{sub}}_2\right],
\end{eqnarray}
in which each bracket is separately finite.%
\footnote{As an independent cross check the phase-space slicing
method has been applied to the subchannel $q\bar q\to t\bar tH$, which
dominates at the Tevatron. The results obtained by the slicing and the
subtraction techniques are in mutual agreement.}

\vspace*{0.5em}
\noindent
{\bf 3.} The results for the \underline{Tevatron} are displayed in
Figs.~\ref{fig:tev}(a,b).  For a Higgs mass between 100 and $150\GeV$,
the cross section varies between about 10 and $1\fb$, the central
value $\mu\to\mu_0=(2m_t+M_H)/2$ chosen for the renormalization and
factorization scales. In NLO the theoretical prediction is remarkably
stable with very little variation for $\mu$ between $\sim\mu_0/3$ and
$\sim 3\mu_0$, in contrast with the Born approximation for which the
production cross section changes by more than a factor 2 within the
same interval. Although at the Tevatron the cross section is strongly
dominated by the $q\bar q$ annihilation channel for scales
$\mu\sim\mu_0$, the proper study of the scale dependence requires
inclusion of the $gg$, $gq$, and $g\bar q$ channels. If $\mu$ is
chosen too low, large logarithmic corrections spoil the convergence of
perturbation theory, and the NLO cross section would even turn negative
for $\mu \lsim \mu_0/5$.

As apparent from Fig.~\ref{fig:tev}(b), the $K$ factor, $K=\sigma_{\rm
NLO}/\sigma_{\rm LO}$ with all quantities calculated consistently in
lowest and next-to-leading order, varies from $\sim 0.8$ at the
central scale $\mu=\mu_0$ to $\sim 1.0$ at the threshold scale
$\mu=2\mu_0$.  The small $K$ factor can be understood intuitively in
the fragmentation picture proposed in Ref.\cite{Dawson:1998im}.  The
average CM energy $\langle\sqrt{\hat s}\rangle$ for the subprocess
$q\bar q\to t\bar tH$ at the Tevatron is about $650\GeV$, i.e.\
sufficiently above the threshold region, so that the EHA of
Ref.\cite{Dawson:1998im} can be used at least at a qualitative level,
as confirmed earlier for $e^+ e^-\to t\bar tH$
\cite{Djouadi:1992tk}.
For $M_H^2\ll m_t^2\ll\langle\sqrt{\hat s}\rangle^2$, the probability
for the hadronic process is decomposed into the product of
probabilities for $t\bar t$ production and subsequent fragmentation
$t\to t+H$. As a result, the relative QCD corrections take the form
$\delta = \delta[p\bar p\to t\bar t\,]+\delta[t\to t+H]$.  With
$\delta[p\bar p\to q\bar q \to t\bar t\,]\sim
-\alpha_s/2\pi$~\cite{Nason:1988xz}
and $\delta[t\to t+H]\sim -4\alpha_s/\pi$
for small energies of the Higgs boson, the sum of the two terms
$\delta  \sim -9 \alpha_s / 2 \pi$ is negative and the $K$ factor is
predicted below unity in this limit. Integrating over the entire
Higgs spectrum, the numerical evaluation yields $K_{\mathrm{EHA}}\sim 0.7$,
which is nicely compatible with the result $K\sim 0.8$ of the full
${\cal O}(\alpha_s)$ calculation.

Near threshold, $\sqrt{\hat s}\gsim2m_t+M_H$, the QCD corrections
are enhanced by Coulombic gluon exchange between the top and antitop-quark
in the final state. This Sommerfeld rescattering correction
\cite{Sommerfeld} increases inversely proportional to the maximum $t/\bar t$
velocity
$\hat\beta_t^{\mathrm{max}}\sim\sqrt{(\sqrt{\hat s}-M_H)^2-4 m_t^2}/2m_t$
in the $t\bar t$ CM frame: $\delta_{\mathrm{Coul}}(\hat s) =
C\cdot \pi\alpha_s/2\cdot \langle 1/\hat\beta_t \rangle =
C\cdot 8\alpha_s/3\hat\beta_t^{\mathrm{max}}$.
If the $t\bar t$ pair is generated in a color-singlet state, the
quark and anti-quark attract each other, and with $C_1=+4/3$ the correction
is positive. This leads to a strong enhancement of the
$e^+e^-\to t\bar t H$ annihilation cross section near threshold
\cite{Djouadi:1992tk}. By contrast, if the $t\bar t$ pair is generated in a
color-octet state, the force is repulsive, and with $C_8=-1/6$ the
correction is negative and relatively small.  This applies to the dominant
channel at the Tevatron, $q\bar q\to t\bar tH$, which is mediated by
$s$-channel color-octet gluon exchange. As a consequence, the destructive
Coulomb interference term amplifies the reduction of the cross section.
%at higher CM energies.
\begin{figure}[hbt]
\centerline{\unitlength 0.6cm
\begin{picture}(15.5,20.0)
\put(0.0,3.0){\includegraphics{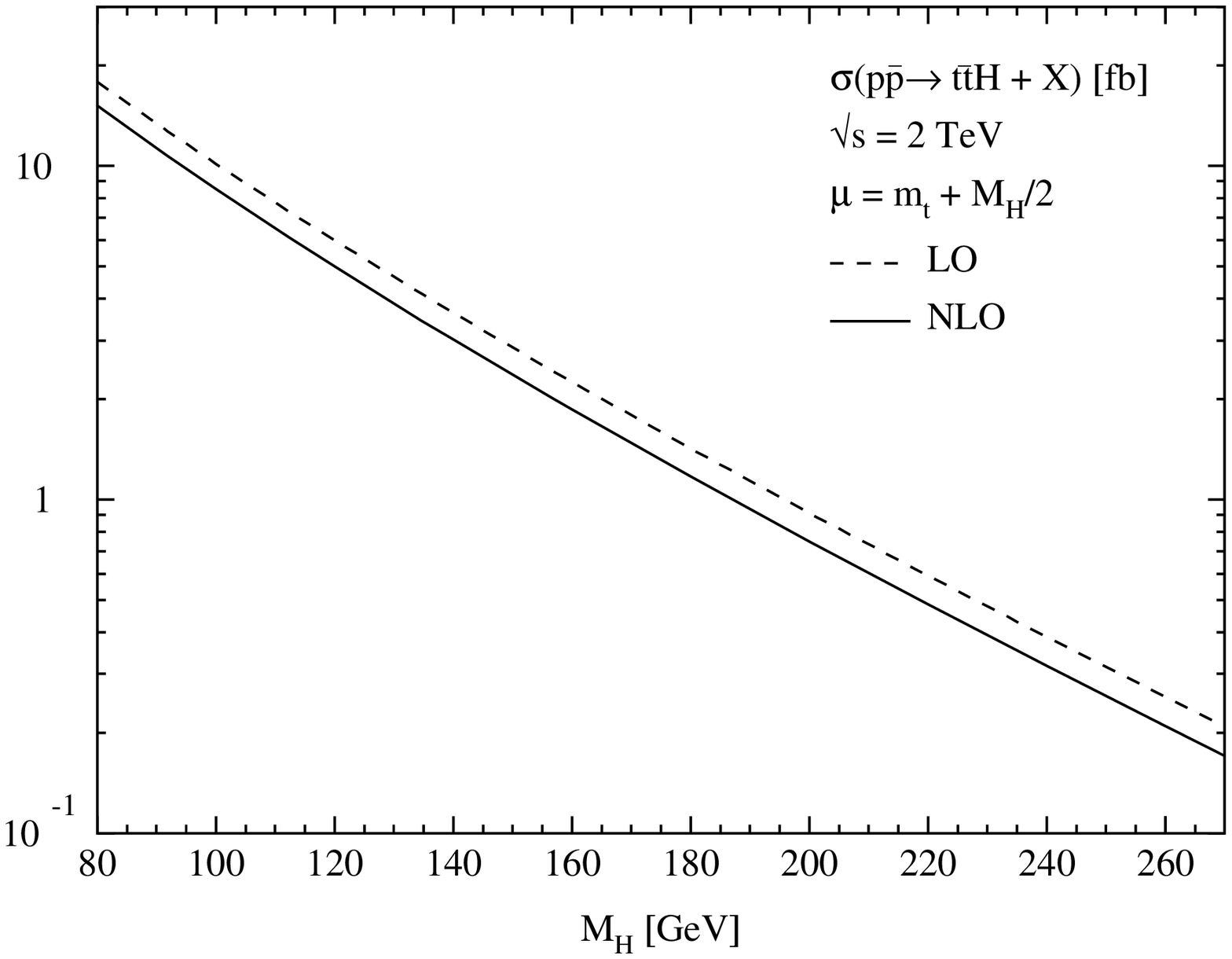}}
\put(0.0,-8.5){\includegraphics{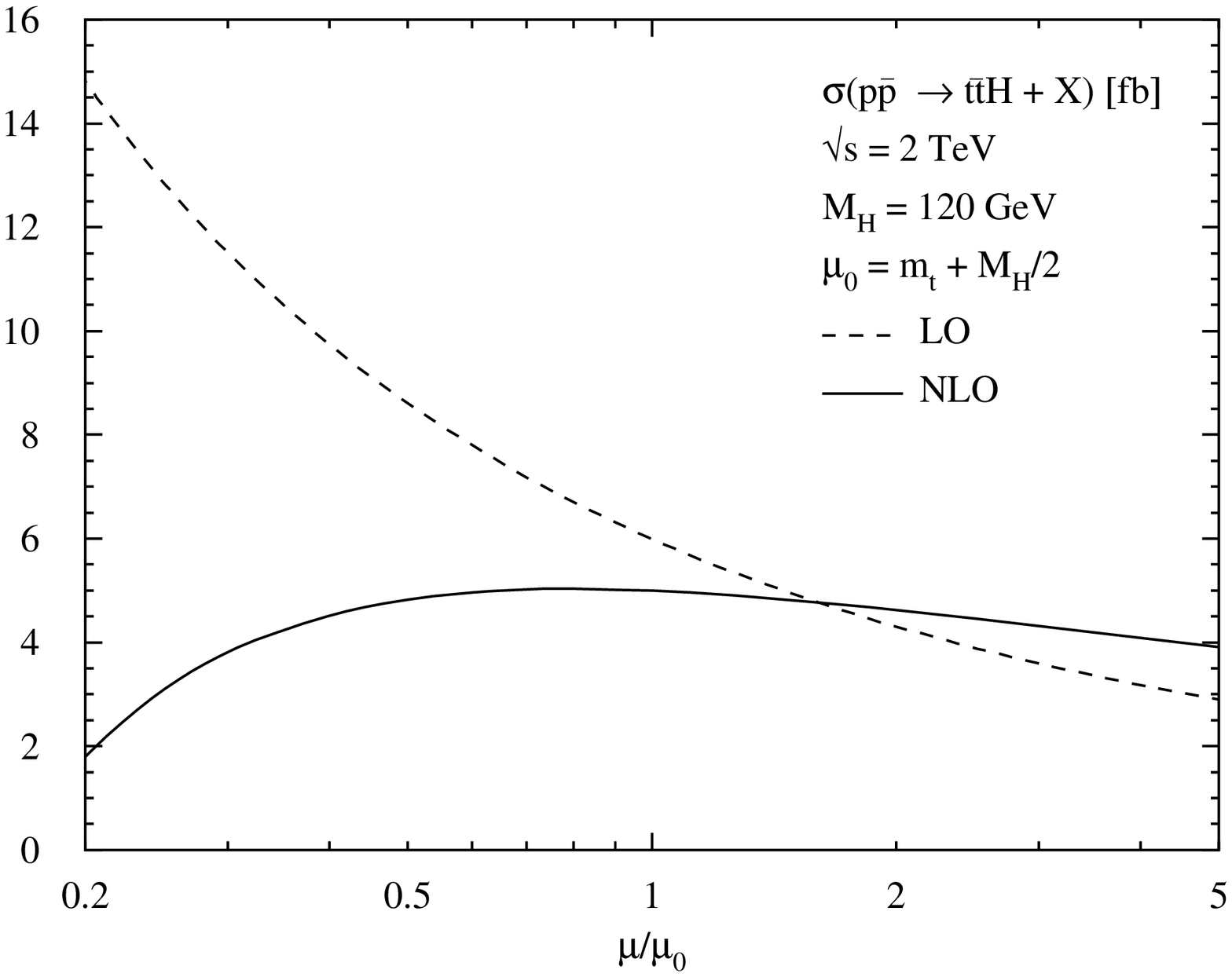}}
\put( .5,19.5){{\bf (a)}}
\put( .5, 8.5){{\bf (b)}}
\end{picture}}
\vspace*{5.0em}
\caption{(a) The cross section for $p\bar p\to t\bar tH\;+\;X$ at the Tevatron
in LO and NLO approximation, with the renormalization and
factorization scales set to $\mu=m_t+M_H/2$; (b) variation of the
cross section with the renormalization and factorization scales for a
fixed Higgs-boson mass $M_H=120\GeV$.}
\label{fig:tev}
\end{figure}

\vspace*{0.5em}
\noindent
{\bf 4.} The improvement of the prediction for the cross section at the
\underline{LHC}, Figs.~\ref{fig:lhc}(a,b), is similarly striking. However,
the gluon initial states give rise to increased gluon radiative corrections
[which will be improved by resummation techniques in the future].
For the central renormalization/factorization
scale $\mu_0$ we obtain $K\sim 1.2$, increasing to $\sim 1.4$ at the
threshold value $\mu=2\mu_0$. These values are nearly
independent of $M_H$ in the relevant Higgs mass range.

The $K$ factor at the LHC can be estimated in the fragmentation
picture \cite{Dawson:1998im}, since the average subenergy
$\langle\sqrt{\hat s}\rangle\sim 830\GeV$ is relatively high at the
LHC.  With the dominant $gg$ production channel, the sum of
$\delta[pp\to gg \to t\bar t\,]\sim +11\alpha_s/\pi$ and
$\delta[t\to t+H]\sim -4\alpha_s/\pi$
for the QCD corrections come now with opposite signs, but the
positive correction to the $t\bar{t}$ production in the $gg$ channel
more than compensates the negative correction to the
fragmentation, leading  finally to $\delta \sim +7 \alpha_s / \pi$.
Taking into account the different renormalization and
factorization scales, the estimate for the $K$ factor in
Ref.~\cite{Dawson:1998im} is recognized compatible at the qualitative
level with the full NLO result.

\begin{figure}[hbt]
\centerline{\unitlength 0.6cm
\begin{picture}(15.5,20.0)
\put(0.0,3.0){\includegraphics{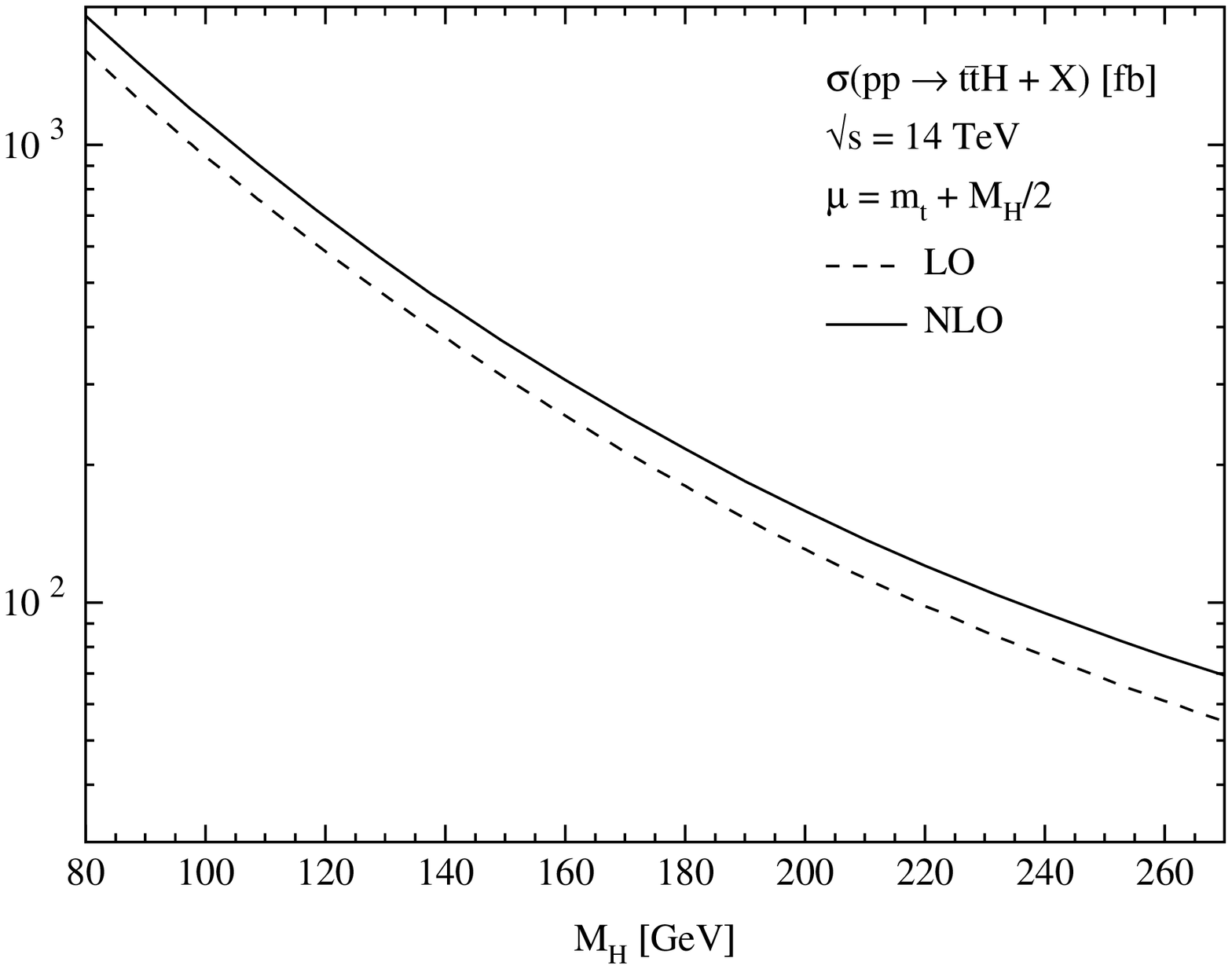}}
\put(0.0,-8.5){\includegraphics{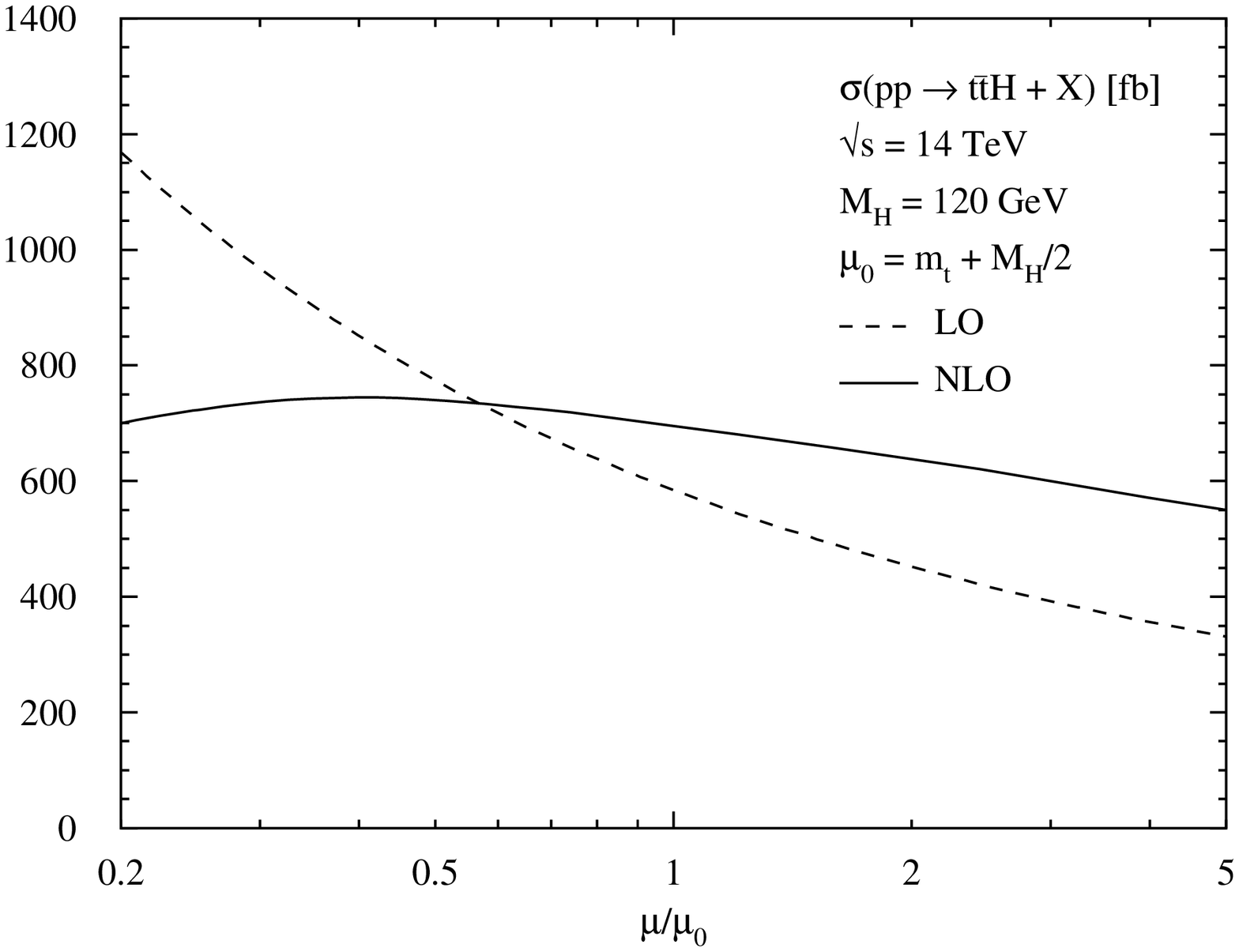}}
\put( .5,19.5){{\bf (a)}}
\put( .5, 8.5){{\bf (b)}}
\end{picture}}
\vspace*{5.0em}
\caption{Analysis of $pp\to t\bar tH\;+\;X$ at the LHC;
(a) production cross section, and (b) renormalization/factorization-scale
dependence (parameters as specified in the previous figure).}
\label{fig:lhc}
\end{figure}

\vspace*{0.5em}
\noindent
{\bf 5.} Summary. The strong scale dependence of the Born cross
sections in the reactions $pp/p\bar p\to t\bar tH$, which provide
important search channels for the SM Higgs boson, requires the
improvement by NLO QCD corrections. In agreement with a qualitative
fragmentation picture, the $K$ factor at the Tevatron is slightly
below unity, i.e.\ varying between $\sim 0.8$ and $\sim 1.0$ for
renormalization and factorization scales $\mu=\mu_0$ and $2\mu_0$,
with $2\mu_0$ denoting the threshold CM energy of the parton
subprocesses.  Similarly, the $K$ factor varies between $\sim 1.2$
and $\sim 1.4$ for the same scale at the LHC. Most important, in
contrast to the Born approximation, the predictions for the cross
sections including NLO QCD corrections are stable when the
renormalization and factorization scales are varied so that the
improved cross sections can serve as a solid base for experimental
analyses at the Tevatron and the LHC.
\\[3.5em]
{\bf Acknowledgements}
\\[0.5em]
The analysis of the subprocess $q\bar q\to t\bar tH$ at the Tevatron
has been compared with the parallel calculation of
Ref.\cite{Dawson:2001}; the results are in agreement. We thank
Sally Dawson and Laura Reina for a pleasant cooperation.  M.K.\ would
like to thank Keith Ellis for discussions and the DESY Theory Group for
hospitality. This work has been supported by the European Union under
contract HPRN-CT-2000-00149.

\end{narrowtext}

\begin{references}

%\cite{Higgs:1964ia}
\bibitem{Higgs:1964ia}
P.~W.~Higgs,
%``Broken Symmetries, Massless Particles And Gauge Fields,''
Phys.\ Lett.\  {\bf 12}, 132 (1964);
%%CITATION = PHLTA,12,132;%%
%\cite{Higgs:1964pj}
%\bibitem{Higgs:1964pj}
%P.~W.~Higgs,
%``Broken Symmetries And The Masses Of Gauge Bosons,''
Phys.\ Rev.\ Lett.\  {\bf 13}, 508 (1964)
%%CITATION = PRLTA,13,508;%%
%\cite{Higgs:1966ev}
%\bibitem{Higgs:1966ev}
%P.~W.~Higgs,
%``Spontaneous Symmetry Breakdown Without Massless Bosons,''
and Phys.\ Rev.\  {\bf 145}, 1156 (1966);
%%CITATION = PHRVA,145,1156;%%
%\cite{Englert:1964et}
%\bibitem{Englert:1964et}
F.~Englert and R.~Brout,
%``Broken Symmetry And The Mass Of Gauge Vector Mesons,''
Phys.\ Rev.\ Lett.\  {\bf 13}, 321 (1964);
%%CITATION = PRLTA,13,321;%%
%\cite{Guralnik:1964eu}
%\bibitem{Guralnik:1964eu}
G.~S.~Guralnik, C.~R.~Hagen and T.~W.~Kibble,
%``Global Conservation Laws And Massless Particles,''
Phys.\ Rev.\ Lett.\  {\bf 13}, 585 (1964).
%%CITATION = PRLTA,13,585;%%

%\cite{Carena:2000yx}
\bibitem{Carena:2000yx}
M.~Carena, J.S.~Conway, H.E.~Haber and J.D.~Hobbs {\it et al.},
%``Report of the Tevatron Higgs working group,''
hep-ph/0010338.
%%CITATION = HEP-PH 0010338;%%

%\cite{atlas_cms_tdrs}
\bibitem{atlas_cms_tdrs}
ATLAS Collaboration, Technical Design Report, CERN--LHCC 99--14
 (May 1999);
%\cite{atlas_cms_tdrs}
%\bibitem{atlas_cms_tdrs}
CMS Collaboration, Technical Proposal, CERN--LHCC 94--38 (Dec.~1994).

%\cite{:2001xv}
\bibitem{:2001xv}
[The LEP Collaborations],
%``A combination of preliminary electroweak measurements and constraints  on the standard model,''
hep-ex/0103048;
%%CITATION = HEP-EX 0103048;%%
T.~Ka\-wa\-mo\-to, Talk at the {\it XXXVIth Rencontres de Moriond-2001,
QCD and hadronic interactions,} March 2001, Les Arcs,
hep-ex/0105032;
E.~Tournefier, Talk at the {\it XXXVIth Rencontres de Moriond-2001,
Electroweak interactions and unified theories,} March 2001, Les Arcs,
hep-ex/0105091;
%\cite{Erler:2000tn}
%\bibitem{Erler:2000tn}
J.~Erler,
%``Fundamental parameters from precision tests,''
hep-ph/0102143.
%%CITATION = HEP-PH 0102143;%%

%\cite{Chanowitz:2001bv}
\bibitem{Chanowitz:2001bv}
M.~S.~Chanowitz,
%``The Z $\to$ anti-b b decay asymmetry: Lose-lose for the standard model,''
hep-ph/0104024;
%%CITATION = HEP-PH 0104024;%%
%\cite{Bagger:2000te}
%\bibitem{Bagger:2000te}
J.~A.~Bagger, A.~F.~Falk and M.~Swartz,
%``Precision observables and electroweak theories,''
Phys.\ Rev.\ Lett.\  {\bf 84}, 1385 (2000)
[hep-ph/9908327];
%%CITATION = HEP-PH 9908327;%%
%\cite{Peskin:2001rw}
%\bibitem{Peskin:2001rw}
M.~E.~Peskin and J.~D.~Wells,
%``How can a heavy Higgs boson be consistent with the precision  electroweak measurements?,''
hep-ph/0101342.
%%CITATION = HEP-PH 0101342;%%

%\cite{Kunszt:1984ri}
\bibitem{Kunszt:1984ri}
Z.~Kunszt,
%``Associated Production Of Heavy Higgs Boson With Top Quarks,''
Nucl.\ Phys.\ B {\bf 247}, 339 (1984);
%%CITATION = NUPHA,B247,339;%%
%\cite{Marciano:1991qq}
%\bibitem{Marciano:1991qq}
W.~J.~Marciano and F.~E.~Paige,
%``Associated production of Higgs bosons with t anti-t pairs,''
Phys.\ Rev.\ Lett.\  {\bf 66}, 2433 (1991);
%%CITATION = PRLTA,66,2433;%%
%\cite{Gunion:1991kg}
%\bibitem{Gunion:1991kg}
J.~F.~Gunion,
%``Associated top anti-top Higgs production as a large source of W H events: Implications for Higgs detection in the lepton neutrino gamma gamma final state,''
Phys.\ Lett.\ B {\bf 261}, 510 (1991).
%%CITATION = PHLTA,B261,510;%%

%\cite{Goldstein:2001bp}
\bibitem{Goldstein:2001bp}
J.~Goldstein, C.~S.~Hill, J.~Incandela, S.~Parke, D.~Rainwater and D.~Stuart,
%``p anti-p $\to$ t anti-t H: A discovery mode for the Higgs boson at the  Tevatron,''
Phys.\ Rev.\ Lett.\  {\bf 86}, 1694 (2001)
[hep-ph/0006311].
%%CITATION = HEP-PH 0006311;%%

%\cite{Drollinger:2001xm}
\bibitem{Drollinger:2001xm}
M.~Beneke {\it et al.}, in proceedings of the workshop on "Standard
Model Physics (and more) at the LHC'', eds. G~Altarelli and
M.L.~Mangano, Geneva 2000, CERN 2000-04 [hep-ph/0003033];
V.~Drollinger,
%``Finding H0 $\to$ b anti-b at the LHC,''
hep-ex/0105017.
%%CITATION = HEP-EX 0105017;%%

%\cite{Djouadi:1992tk}
\bibitem{Djouadi:1992tk}
A.~Djouadi, J.~Kalinowski and P.~M.~Zerwas,
%``Higgs radiation off top quarks in high-energy e+ e- colliders,''
Z.\ Phys.\ C {\bf 54}, 255 (1992);
%%CITATION = ZEPYA,C54,255;%%
%\cite{Dittmaier:1998dz}
%\bibitem{Dittmaier:1998dz}
S.~Dittmaier, M.~Kr\"amer, Y.~Liao, M.~Spira and P.~M.~Zerwas,
%``Higgs radiation off top quarks in e+ e- collisions,''
Phys.\ Lett.\ B {\bf 441}, 383 (1998)
[hep-ph/9808433];
%%CITATION = HEP-PH 9808433;%%
%\cite{Dawson:1999ej}
%\bibitem{Dawson:1999ej}
S.~Dawson and L.~Reina,
%``{QCD} corrections to associated Higgs boson-heavy quark production,''
Phys.\ Rev.\ D {\bf 59}, 054012 (1999)
[hep-ph/9808443].
%%CITATION = HEP-PH 9808443;%%

%\cite{Dawson:1998im}
\bibitem{Dawson:1998im}
S.~Dawson and L.~Reina,
%``QCD corrections to associated Higgs boson production,''
Phys.\ Rev.\ D {\bf 57}, 5851 (1998)
[hep-ph/9712400].
%%CITATION = HEP-PH 9712400;%%

%\cite{Lai:1997mg}
\bibitem{Lai:1997mg}
H.~L.~Lai {\it et al.},
%``Improved parton distributions from global analysis of recent deep  inelastic scattering and inclusive jet data,''
Phys.\ Rev.\ D {\bf 55}, 1280 (1997)
[hep-ph/9606399].
%%CITATION = HEP-PH 9606399;%%

%\cite{Melrose:1965kb}
\bibitem{Melrose:1965kb}
D.~B.~Melrose,
%``Reduction Of Feynman Diagrams,''
Nuovo Cim.\  {\bf 40}, 181 (1965).
%%CITATION = NUCIA,40,181;%%

%\cite{'tHooft:1979xw}
\bibitem{'tHooft:1979xw}
G.~'t Hooft and M.~Veltman,
%``Scalar One Loop Integrals,''
Nucl.\ Phys.\ B {\bf 153}, 365 (1979);
%%CITATION = NUPHA,B153,365;%%
%\cite{Passarino:1979jh}
%\bibitem{Passarino:1979jh}
G.~Passarino and M.~Veltman,
%``One Loop Corrections For E+ E- Annihilation Into Mu+ Mu- In The Weinberg Model,''
Nucl.\ Phys.\ B {\bf 160}, 151 (1979);
%%CITATION = NUPHA,B160,151;%%
%\cite{Beenakker:1990jr}
%\bibitem{Beenakker:1990jr}
W.~Beenakker and A.~Denner,
%``Infrared Divergent Scalar Box Integrals With Applications In The Electroweak Standard Model,''
Nucl.\ Phys.\ B {\bf 338}, 349 (1990);
%%CITATION = NUPHA,B338,349;%%
%\cite{Denner:1991qq}
%\bibitem{Denner:1991qq}
A.~Denner, U.~Nierste and R.~Scharf,
%``A Compact expression for the scalar one loop four point function,''
Nucl.\ Phys.\ B {\bf 367}, 637 (1991).
%%CITATION = NUPHA,B367,637;%%

%\cite{Catani:1996jh}
\bibitem{Catani:1996jh}
S.~Catani and M.~H.~Seymour,
%``The Dipole Formalism for the Calculation of QCD Jet Cross Sections at Next-to-Leading Order,''
Phys.\ Lett.\ B {\bf 378} (1996) 287
[hep-ph/9602277] and
%%CITATION = HEP-PH 9602277;%%
%\cite{Catani:1997vz}
%\bibitem{Catani:1997vz}
%S.~Catani and M.~H.~Seymour,
%``A general algorithm for calculating jet cross sections in NLO QCD,''
Nucl.\ Phys.\ B {\bf 485} (1997) 291
[Erratum - {\it ibid}.\ B {\bf 510} (1997) 291]
[hep-ph/9605323].
%%CITATION = HEP-PH 9605323;%%

\bibitem{Catani:2001}
S.~Catani, S.~Dittmaier, M.H.~Seymour and Z.~Tr\'ocs\'anyi, in preparation.

%\cite{Dittmaier:2000mb}
\bibitem{Dittmaier:2000mb}
S.~Dittmaier,
%``A general approach to photon radiation off fermions,''
Nucl.\ Phys.\ B {\bf 565} (2000) 69
[hep-ph/9904440];
%%CITATION = HEP-PH 9904440;%%
%\cite{Catani:2001ef}
%\bibitem{Catani:2001ef}
S.~Catani, S.~Dittmaier and Z.~Tr\'ocs\'anyi,
%``One-loop singular behavior of QCD and SUSY QCD amplitudes with  massive partons,''
Phys.\ Lett.\ B {\bf 500} (2001) 149
[hep-ph/0011222];
%%CITATION = HEP-PH 0011222;%%
%\cite{Phaf:2001gc}
%\bibitem{Phaf:2001gc}
L.~Phaf and S.~Weinzierl,
%``Dipole formalism with heavy fermions,''
JHEP {\bf 0104}, 006 (2001)
[hep-ph/0102207].
%%CITATION = HEP-PH 0102207;%%

\bibitem{Nason:1988xz}
P.~Nason, S.~Dawson and R.~K.~Ellis,
%``The Total Cross-Section For The Production Of Heavy Quarks In Hadronic Collisions,''
Nucl.\ Phys.\ B {\bf 303} (1988) 607,
%%CITATION = NUPHA,B303,607;%%
and Nucl.\ Phys.\ B {\bf 327} (1989) 49;
%%CITATION = NUPHA,B327,49;%%
%\cite{Beenakker:1989bq}
%\bibitem{Beenakker:1989bq}
W.~Beenakker, H.~Kuijf, W.~L.~van Neerven and J.~Smith,
%``QCD Corrections To Heavy Quark Production In P Anti-P Collisions,''
Phys.\ Rev.\ D {\bf 40} (1989) 54;
%%CITATION = PHRVA,D40,54;%%
%\cite{Beenakker:1991ma}
%\bibitem{Beenakker:1991ma}
W.~Beenakker, W.~L.~van Neerven, R.~Meng, G.~A.~Schuler and J.~Smith,
%``QCD corrections to heavy quark production in hadron-hadron collisions,''
Nucl.\ Phys.\ B {\bf 351} (1991) 507.
%%CITATION = NUPHA,B351,507;%%

\bibitem{Sommerfeld}
A.~Sommerfeld, in ``Atombau und Spektrallinien'', Vol.~II, p.~457
(F.~Vieweg und Sohn, Braunschweig 1939).

\bibitem{Dawson:2001}
L.~Reina and S.~Dawson, FSU-HEP-2001-0601; L.~Reina, S.~Dawson and
D.~Wackeroth, FSU-HEP-2001-0602.

%\end{thebibliography}
\end{references}
\end{document}